\documentclass[journal=jpclcd,manuscript=letter]{achemso}

\usepackage[version=3]{mhchem} 

\usepackage{hyperref}

\author{Michele Invernizzi}
\affiliation{Department of Physics, ETH Zurich, c/o Universit\`{a} della Svizzera italiana, Via Giuseppe Buffi 13, 6900, Lugano, Switzerland}
\affiliation{Facolt\`{a} di Informatica, Institute of Computational Science,
National Center for Computational Design and Discovery of Novel Materials (MARVEL), Universit\`{a} della Svizzera italiana, Via Giuseppe Buffi 13, 6900 Lugano, Switzerland}
\author{Michele Parrinello}
\email{parrinello@phys.chem.ethz.ch}
\affiliation{Department of Chemistry and Applied Biosciences, ETH Zurich, c/o Universit\`{a} della Svizzera italiana, Via Giuseppe Buffi 13, 6900 Lugano, Switzerland,
and Italian Institute of Technology, Via Morego 30, 16163 Genova, Italy}
\affiliation{Facolt\`{a} di Informatica, Institute of Computational Science,
National Center for Computational Design and Discovery of Novel Materials (MARVEL), Universit\`{a} della Svizzera italiana, Via Giuseppe Buffi 13, 6900 Lugano, Switzerland}

\title{Rethinking Metadynamics:\\from bias potentials to probability distributions}

\abbreviations{CV,AUS,FES,MetaD,VES,KDE,OPES}
\keywords{Free Energy, Enhanced Sampling, Rare Events, Importance Sampling}

\begin{document}

\begin{tocentry}
    \includegraphics[width=5cm]{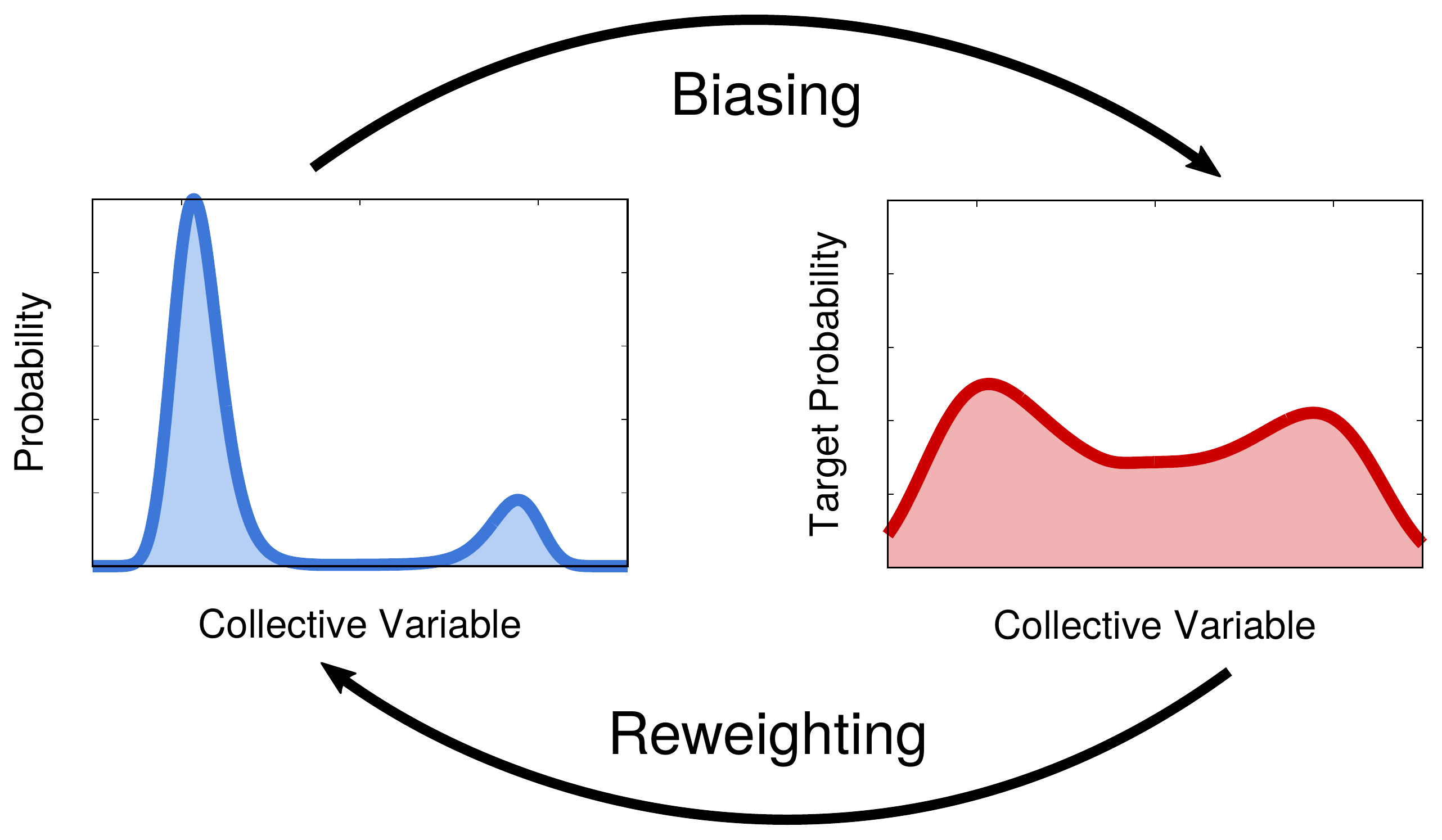}
\end{tocentry}

\begin{abstract}
  Metadynamics is an enhanced sampling method of great popularity, based on the on-the-fly construction of a bias potential that is function of a selected number of collective variables.
  We propose here a change in perspective that shifts the focus from the bias to the probability distribution reconstruction, while keeping some of the key characteristics of metadynamics, such as the flexible on-the-fly adjustments to the free energy estimate.
  The result is an enhanced sampling method that presents a drastic improvement in convergence speed, especially when dealing with suboptimal and/or multidimensional sets of collective variables.
  The method is especially robust and easy to use, in fact it requires only few simple parameters to be set, and it has a straightforward reweighting scheme to recover the statistics of the unbiased ensemble.
  Furthermore it gives more control on the desired exploration of the phase space, since the deposited bias is not allowed to grow indefinitely and it does not push the simulation to uninteresting high free energy regions.
  We demonstrate the performance of the method in a number of representative examples.
\end{abstract}

Enhanced sampling plays a crucial role in modern simulation techniques, and is a very active area of research\cite{Peters2017}.
Of particular historical importance has been the work of Torrie and Valleau\cite{Torrie1977}.
They consider a system with an interaction potential $U(\mathbf{R})$, where $\mathbf{R}$ denotes the atomic coordinates.
Sampling is accelerated by adding a bias potential $V(\mathbf{s})$ that depends on $\mathbf{R}$ via a set of collective variables (CVs), $\mathbf{s}=\mathbf{s}(\mathbf{R})$.
The CVs are chosen so as to describe the modes of the system that are more difficult to sample.
The choice of a proper set of CVs is critical, as it determines the efficiency of the method.
The properties of the unbiased system are then calculated by using a reweighting procedure.
In fact the unbiased probability density $P(\mathbf{s}) = \langle \delta[\mathbf{s}-\mathbf{s}(\mathbf{R})] \rangle \propto \int d \mathbf{R}\, e^{-\beta U(\mathbf{R})} \delta[\mathbf{s}-\mathbf{s}(\mathbf{R})]$ can be written as an average over the biased ensemble:
\begin{equation}\label{E:reweighting}
    P(\mathbf{s}) = \frac{\langle \delta[\mathbf{s}-\mathbf{s}(\mathbf{R})] e^{\beta V(\mathbf{s})}\rangle_V}{\langle e^{\beta V(\mathbf{s})}\rangle_V}\, ,
\end{equation}
where $\beta$ is the inverse temperature.
In this way it is also possible to reconstruct the free energy surface (FES), defined as $F(\mathbf{s})=-\frac{1}{\beta} \log P(\mathbf{s})$.

Since the work of Torrie and Valleau, a large number of CV-based enhanced sampling methods has been proposed.
Among them is metadynamics\cite{Laio2002,Barducci2008} (MetaD) which builds the bias $V(\mathbf{s})$ by adding at fixed intervals repulsive Gaussians centered at the instantaneous point sampled.
At the $n$-th iteration the bias is given by:
\begin{equation}\label{E:metad}
    V_n(\mathbf{s})=\sum_k^n e^{-\beta V_{k-1}(\mathbf{s}_k)/(\gamma-1)}\, G(\mathbf{s},\mathbf{s}_k)\, ,
\end{equation}
where the parameter $\gamma>1$ is called the bias factor, and the Gaussian function is defined as $G(\mathbf{s},\mathbf{s}')=h\exp \left[-\frac{1}{2} (\mathbf{s}-\mathbf{s}')^T\boldsymbol{\Sigma}^{-1} (\mathbf{s}-\mathbf{s}') \right]$, with height $h$ and variance $\boldsymbol{\Sigma}$ set by the user.
Typically only diagonal variances $\Sigma_{ij}=\sigma^2_i\delta_{ij}$ are employed, but more general choices have also been suggested\cite{Branduardi2012}.
It has been proven\cite{Dama2014_convergence} that at convergence there is a simple relation between the bias and the free energy, $V(\mathbf{s})=-(1-1/\gamma) F(\mathbf{s})$, and the sampled ensemble is a smoothed version of the unbiased one, with FES barriers lowered by a factor $\gamma$.

Arguably, a major developments of MetaD has been its well-tempered variant\cite{Barducci2008}.
With only a simple change to the original MetaD equations, it brought many improvements especially regarding the following points.
(1) By damping the bias oscillations it allows for better handling of suboptimal CVs, that is CVs that do not include all the slow modes of the system.
This is a crucial issue, since finding a good CV for a complex system is non-trivial, and even a good CV is usually suboptimal\cite{Pietrucci2017}.
(2) It opens up the possibility of performing reweighting, which is a fundamental aspect of any enhanced sampling method, since it allows to retrieve the unbiased statistics of any quantity of interest.
(3) It gives more control over the regions explored, since the bias does not push the system to extremely high free energy regions.
(4) Thanks to this property, it also improves the handling of multiple CVs, by reducing the volume of CV space that is sampled at convergence.

Despite the success of MetaD, there is certainly room for further improvement.
In fact, over the years many new MetaD variants have been proposed which put particular focus on one of the above mentioned issues\cite{Raiteri2006,Piana2007,Branduardi2012,Dama2014_ttmetad,White2015,Pfaendtner2015}.
Particular attention has been given to reweighting, and many different solutions have been proposed also in very recent years\cite{Bonomi2009,Branduardi2012,Tiwary2015,Mones2016,Donati2018,Salvalaglio2019,Giberti2019}.
With this letter we want to take a step back and propose a new perspective on MetaD, in order to provide a general improvement on all these issues, as it has been the case for well-tempered MetaD.

We start from the observation that in case of suboptimal CVs, the FES estimate obtained via reweighting can converge faster than the bias itself\cite{Invernizzi2017}.
In particular it is more robust, and does not present the strong oscillations typical of MetaD.
Furthermore, a more static bias can help with the reweighting procedure itself, giving rise to a positive feedback loop.
Thus, we develop a method that is based on the reconstruction of the probability distribution via reweighting, and uses this estimate to define the bias potential, rather than directly building it as in Eq.~\ref{E:metad}.

Enhanced sampling based on the probability reconstruction is not a new idea.
It was first proposed by the adaptive umbrella sampling method\cite{Mezei1987}, and later by many others\cite{Maragakis2009,Marsili2006,Dickson2010}.
Typically, in such methods the bias at $n$-th iteration is defined as:
\begin{equation}\label{E:adUS}
    V_n(\mathbf{s})=\frac{1}{\beta}\log \hat{P}_n(\mathbf{s})\, ,
\end{equation}
where $\hat{P}_n(\mathbf{s})$ is an estimate of the probability obtained via a weighted histogram or some more elaborate method\cite{Maragakis2009}, and updated iteratively or on-the-fly\cite{Marsili2006}.
In building our method we will introduce few key differences that come from the long experience with MetaD, which allow us to overcome some of the limitations of previous probability-based methods.

First we would like to introduce explicitly a target distribution $p^{tg}(\mathbf{s})$, that will be sampled once the method reaches convergence.
This can be obtained with the following bias:
\begin{equation}\label{E:converged_bias}
    V(\mathbf{s})=\frac{1}{\beta} \log \frac{P(\mathbf{s})}{p^{tg}(\mathbf{s})}\, .
\end{equation}
In adaptive umbrella sampling the target distribution is uniform, $p^{tg}(\mathbf{s}) \propto 1$, while in MetaD it is the well-tempered distribution, $p^{tg}(\mathbf{s}) \propto [P(\mathbf{s})]^{1/\gamma}$.
It is possible to modify MetaD in order to reach any arbitrary target\cite{White2015}, and in general the concept of a target distribution has proven to be very useful, especially in the contest of variationally enhanced sampling\cite{Valsson2014,Valsson2015,Shaffer2016,Invernizzi2017,Debnath2019,Piaggi2019}.
In the present work we will limit ourselves to a well-tempered target (or a flat target in the $\gamma \rightarrow \infty$ limit) leaving other interesting possibilities for future work.
We notice here that a well-tempered target leads to a more efficient importance sampling compared to the common choice of a flat target, and despite lowering the FES barriers by $\gamma$ instead of flattening them, it generally does not give rise to a slower transition rate between the metastable states.
In fact, in most applications suboptimal CVs are employed and the transition rate is limited by the slow modes not accelerated by $V(\mathbf{s})$ rather than by the small FES barriers left along $\mathbf{s}$ (see also \href{https://arxiv.org/src/1909.07250v4/anc/OPES-SI.pdf}{supporting information - SI})\cite{Invernizzi2019}.

Since we can express the target distribution as a function of the unbiased one, $p^{tg}(\mathbf{s}) \propto [P(\mathbf{s})]^{1/\gamma}$, we only need to estimate $P(\mathbf{s})$ via reweighting in order to calculate the bias.
We build our probability distribution estimate on the fly by periodically depositing Gaussians, similarly to how MetaD builds the bias potential.
This is indeed a common way of reconstructing a probability, known as kernel density estimation (KDE), and we shall draw from the vast literature on the subject\cite{Silverman1998}.
Each new Gaussian is weighted according to the previously deposited bias potential:
\begin{equation}\label{E:iter_prob}
     \Tilde{P}_n(\mathbf{s})=\frac{\sum_k^n w_k G(\mathbf{s},\mathbf{s}_k)}{\sum_k^n w_k}\, ,
\end{equation}
where the weights $w_k$ are given by $w_k=e^{\beta V_{k-1}(\mathbf{s}_k)}$.

We write the estimator in Eq.~\ref{E:iter_prob} with a tilde, $\Tilde{P}_n(\mathbf{s})$, to indicate that it is not properly normalized, and we will take care of the normalization separately.
The $G(\mathbf{s},\mathbf{s}_k)$ are Gaussians as those defined previously for MetaD, with diagonal variance $\Sigma_{ij}=\sigma^2_i\delta_{ij}$ and fixed height $h=\prod_i \left(\sigma_i\sqrt{2\pi}\right)^{-1}$.
Contrary to MetaD, here the height of the Gaussians is not a free parameter, and changing it simply corresponds to changing the overall normalization.

It has been shown\cite{Silverman1998} that in KDE the most relevant parameter is the bandwidth, i.e.~the width of the Gaussians.
A good choice of the bandwidth should depend on the amount of available data: the larger the sampling the smaller the bandwidth.
Thus we choose to shrink the bandwidth as the simulation proceeds according to the popular Silverman's rule of thumb\cite{Silverman1998}.
At $n$-th iteration:
\begin{equation}\label{E:bandwidth}
    \sigma_i^{(n)}=\sigma_i^{(0)} [N_{\text{eff}}^{(n)}(d+2)/4]^{-1/(d+4)}\, ,
\end{equation}
where $\sigma_i^{(0)}$ is the initial standard deviation estimated from a short unbiased simulation, $d$ is the dimensionality of the CV space, and $N_{\text{eff}}^{(n)}=\left(\sum_k^n w_k\right)^2/\sum_k^n w_k^2$ is the effective sample size\cite{Kish1965}.
The KDE literature presents many other promising alternatives for the bandwidth selection, but we leave their study to future investigation.

The number of kernels accumulated during the simulation quickly becomes very large and summing all of them at each time step is prohibitive.
To avoid this problem we adapt to our needs a simple on-the-fly kernel compression algorithm\cite{Sodkomkham2016}, that allows the insertion of new kernels only in newly explored regions, otherwise merges them with existing ones.
In the \href{https://arxiv.org/src/1909.07250v4/anc/OPES-SI.pdf}{supporting information} we discuss this choice in detail, and we show the advantages over the more common approach of storing the bias on a grid\cite{plumed}. 

\begin{figure*}
  \includegraphics[width=\columnwidth]{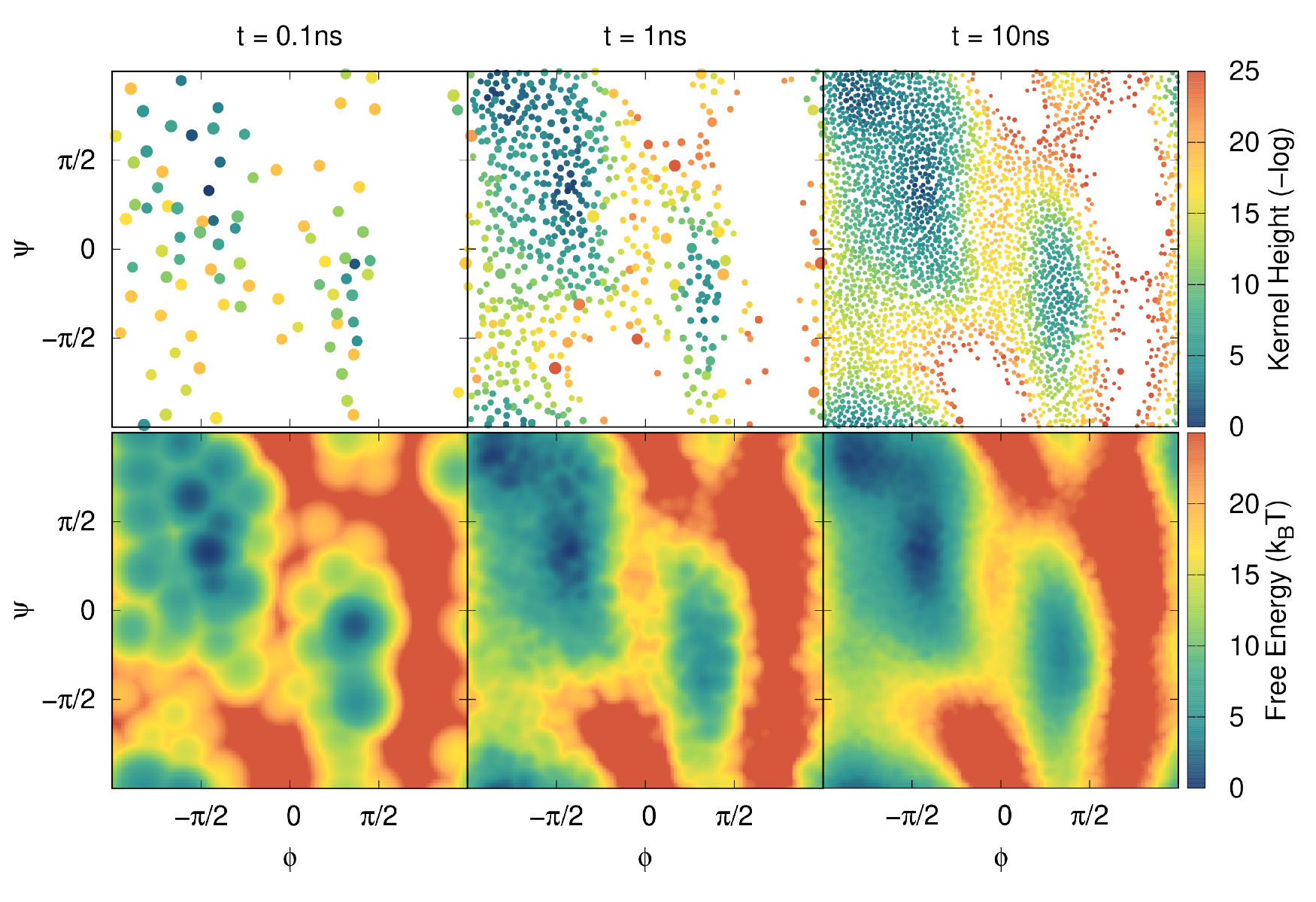}
  \caption{Time evolution of a typical OPES simulation of alanine dipeptide in vacuum, using as CVs the dihedral angles $\phi$ and $\psi$.
    On the left the compressed kernels forming $\Tilde{P}_n(\phi,\psi)$ are shown, with point size indicating the bandwidth, while on the right is the corresponding free energy estimate $F_n(\phi,\psi)=-\frac{1}{\beta}\log \Tilde{P}_n(\phi,\psi)$, shifted to have zero minimum.
    \href{https://arxiv.org/src/1909.07250v4/anc/OPES-SI.pdf}{See SI} for the computational details and a performance comparison with MetaD.}
  \label{F:prob_estimator}
\end{figure*}
In Fig.~\ref{F:prob_estimator} we show how the FES estimate evolves during a typical simulation with our new method.
Our choice of the probability estimator aims at quickly obtaining a coarse representation of the FES and then slowly converging the finer details, and it is one of the key novelties of our method.

We can now discuss the normalization problem.
Any constant overall normalization of the probability estimate $\Tilde{P}_n(\mathbf{s})$ would simply result in a global shift of the bias, thus would not have any influence over the simulation.
However, $\Tilde{P}_n(\mathbf{s})$ should be normalized not with respect to the full CV space, but only over the CV space actually explored up to step $n$, that we call $\Omega_n$.
Thus we introduce the normalization factor
\begin{equation}\label{E:iter_norm}
    Z_n=\frac{1}{|\Omega_n|}\int_{\Omega_n} \Tilde{P}_n(\mathbf{s})\, d \mathbf{s}\, ,
\end{equation}
that will change over time, as the system explores new regions of the CV space, and it will have an impact in the biasing scheme.
This impact becomes particularly relevant in CV spaces of dimension $d \gg 1$, since the volume explored $\Omega_n$ grows with a power of $d$.
Adding such a normalization, together with the chosen probability estimator, helps us overcome the limitations in exploration speed that has affected some of previously proposed on-the-fly probability based methods\cite{Fort2017}.
To estimate $Z_n$ we take advantage of our compressed kernels representation, and substitute the integral in Eq.~\ref{E:iter_norm} with a sum over the positions of the compressed kernels (\href{https://arxiv.org/src/1909.07250v4/anc/OPES-SI.pdf}{see SI}).

Finally we can explicitly write the bias at the $n$-th step as:
\begin{equation}\label{E:iter_bias}
    V_n(\mathbf{s})=(1-1/\gamma)\frac{1}{\beta} \log \left( \frac{\Tilde{P}_n(\mathbf{s})}{Z_n}+\epsilon \right)\, ,
\end{equation}
where $\epsilon \ll 1$ can be seen as a regularization term that ensures the argument of the logarithm is always greater than zero.
We notice that the addition of this term is not merely a technicality to solve a numerical issue, but rather it allows one to set a limit to the bias, thus providing a better control over the desired exploration.
It can be chosen to be $\epsilon=e^{-\beta \Delta E/(1-1/\gamma)}$, where $\Delta E$ is the height of the free energy barriers one wishes to overcome during the enhanced sampling (see also SI).
We could have obtained the same effect of controlling the exploration by properly modifying the target distribution $p^{tg}(\mathbf{s})$, but we believe that introducing $\epsilon$ as a separate term makes for cleaner equations.
By comparing Eq.~\ref{E:iter_bias} and Eq.~\ref{E:adUS} it should be clear that our method distinguishes itself from previous adaptive umbrella sampling methods not only for the employed probability estimator, but also for some other novel key components.

An important feature of our method is that it allows for a simple and straightforward reweighting scheme.
In fact reweighting can be performed in the usual umbrella sampling way (Eq.~\ref{E:reweighting}), without the need for further post processing analysis.
The method has a rapid initial exploration phase, after which a quasi-static regime is reached, but it is by construction robust with respect to the initial non-adiabatic part of the trajectory, thus the reweighting can in practice be performed without cropping out the initial transient (\href{https://arxiv.org/src/1909.07250v4/anc/OPES-SI.pdf}{see SI}).

We implemented the new method, called on-the-fly probability enhanced sampling (OPES), in the enhanced sampling library PLUMED\cite{plumed} and tested it on a variety of different systems.
Here we only provide a quick overview of these tests, but the full results are presented in detail in the \href{https://arxiv.org/src/1909.07250v4/anc/OPES-SI.pdf}{supporting information}.
The code and all the files needed to reproduce the simulations are openly available on the PLUMED-NEST website\cite{Plumed2019}, as plumID:19.068\,.

A full comparison of different enhanced sampling methods is a non-trivial task, and is not the goal of the present paper.
However, in order to give a better idea of our method we compare it with standard well-tempered metadynamics, whose performances might be already familiar to many readers.

We want to test the methods in an agnostic fashion, using very standard input parameters rather than running multiple different simulations and choosing the best performing ones.
One strength of OPES is that it is very simple to setup and needs just three main parameters: the pace at which the bias is updated, the initial bandwidth of the Gaussian kernels, and the approximate height of the barriers one wishes to cross.
In our tests we always keep the deposition pace equal to the one used in MetaD, typically 500 simulation steps.
The initial bandwidth is simply chosen to be equal to the smaller standard deviation of the CVs in the minima, which can be measured in a short unbiased run.
The choice of the barrier parameter requires a minimal knowledge of the system, but only a vague idea is usually enough.
This barrier parameter is used to set to a reasonable default both the regularization factor $\epsilon$ and the bias factor $\gamma$ (\href{https://arxiv.org/src/1909.07250v4/anc/OPES-SI.pdf}{see SI}).
It should be noticed that the choice of $\gamma$ is not as critical as in MetaD, since here it does not directly influence the convergence speed, but only the shape of the target distribution.
In fact in OPES the limit $\gamma \rightarrow \infty$ is not problematic, as in MetaD\cite{Lelievre2010}, and OPES can converge also to the flat target distribution.
In our tests we always used the same value of $\gamma$ for both OPES and MetaD.

\begin{figure}
  \includegraphics[width=0.5\columnwidth]{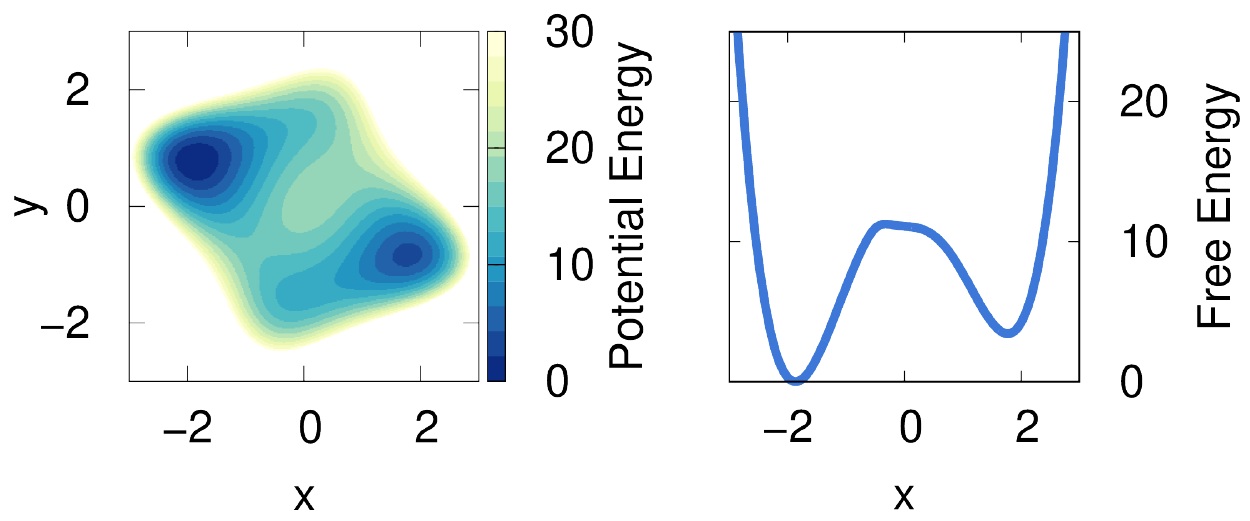}
  \caption{The potential energy of the suboptimal double well 2D model system, and its free energy along the $x$ coordinate.}
  \label{F:model}
\end{figure}
\begin{figure}
  \includegraphics[width=0.5\columnwidth]{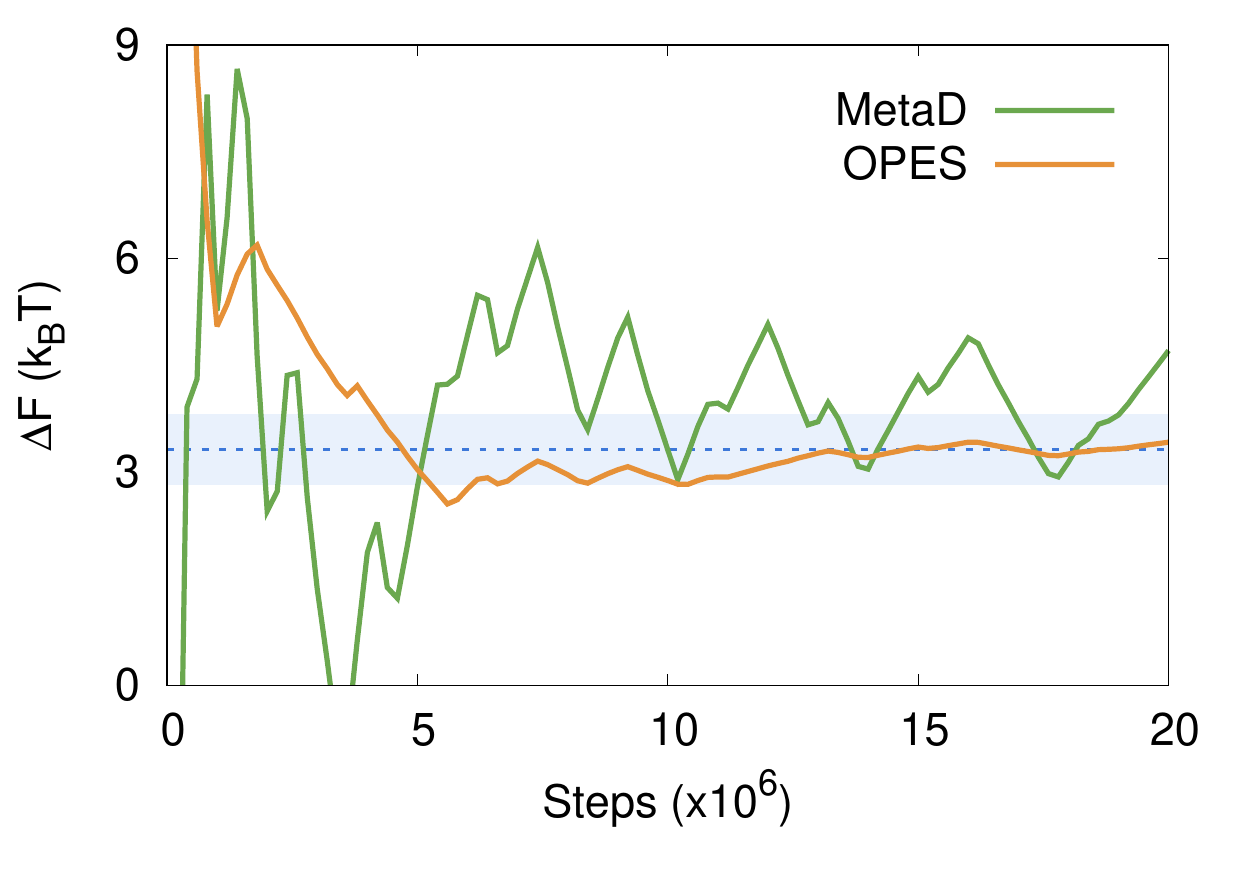}
  \caption{A typical time evolution of the free energy difference between the two basins of the model in Fig.~\ref{F:model}.
    We run the same simulation with 100 different initial conditions, and in the SI we show the average and uncertainty obtained from these estimates.
    The reference blue stripe is $1 k_BT$ thick.}
  \label{F:deltaF}
\end{figure}

In order to test the convergence speed in case of suboptimal CVs, we consider a Langevin dynamics on a 2D model potential\cite{Invernizzi2019}, Fig.~\ref{F:model}, and bias only the $x$ coordinate.
In Fig.~\ref{F:deltaF} we compare MetaD and OPES, by plotting the estimate of the free energy difference between the two basins as a function of time. 
Such estimates are obtained directly from the applied bias, as $F_n(x)=V_n(x)/(1/\gamma-1)$. 
From Fig.~\ref{F:deltaF} one can see that as OPES converges, it does not present the strong bias oscillations typical of MetaD.

\begin{figure}
  \includegraphics[width=0.5\columnwidth]{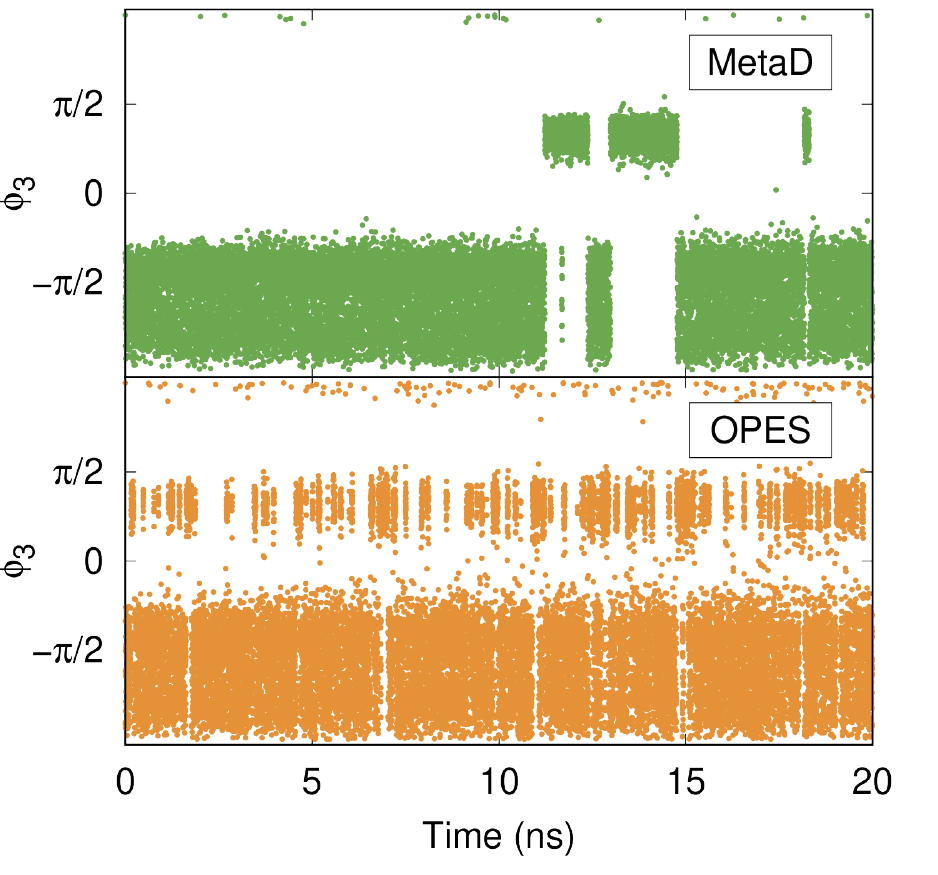}
  \caption{The $\phi_3$ trajectory for alanine tetrapeptide in vacuum, obtained by biasing all the 6 dihedral angles ($\phi_1, \phi_2, \phi_3, \psi_1, \psi_2, \psi_3$), using MetaD and OPES, with the same input parameters as the standard ones used for alanine dipeptide.
        We show the $\phi_3$ angle because it is the hardest one to be sampled, but in the SI all of them are presented, together with the reconstructed FES.}
  \label{F:colvar}
\end{figure}
We also run the typical benchmark system for novel enhanced sampling methods, alanine dipeptide, biasing the two dihedral angles $\phi$ and $\psi$.
Results are in Fig.~\ref{F:prob_estimator} and in the SI.

As an example of multidimensional bias, we run simulations of alanine tetrapeptide in vacuum, biasing all its six dihedral angles.
In Fig.\ref{F:colvar} we show how OPES is able to explore this high dimensional CV space much more efficiently than MetaD.
It is important to notice that we use the same input parameters for alanine dipeptide and alanine tetrapeptide.
This should be a reasonable choice, since the two systems are very similar from the point of view of the physics involved, the main difference being the increased dimensionality of the CV space.
However, many CV-based enhanced sampling methods would require further tuning or a different set of inputs in order to perform well on both systems.
This is not the case for OPES, thanks to its robustness to the choice of the input parameters.

In conclusion, with this letter we present a new enhanced sampling method, OPES, based on an on-the-fly reconstruction of the probability distribution.
It performs such reconstruction via a weighted kernel density estimation (Eq.~\ref{E:iter_prob}) with an on-the-fly compression algorithm that allows it to be most effective, starting from a coarse-grained guess of the free energy surface and then converging the finer details.
Thanks to this strategy and to the introduction of a normalization over the explored CV space (Eq.~\ref{E:iter_norm}), the method provides an extremely fast exploration, also in relatively high dimensions.
Another peculiarity of the method is the presence of an upper limit to the applied bias (Eq.\ref{E:iter_bias}), which can be useful to avoid sampling unphysical states.
Most importantly the proposed method requires few simple and robust input parameters, has very good convergence performance and presents a straightforward reweighting scheme.
Finally, we considered here only the case of a well-tempered target, but OPES provides a general framework in which also different targets can be considered.

We believe this new method can become a handy tool in addressing enhanced sampling problems, and has the potential of opening to further interesting developments.

\begin{acknowledgement}

This research was supported by the NCCR MARVEL, funded by the Swiss National Science Foundation, and European Union Grant No.~ERC-2014-AdG-670227/VARMET. 
Calculations were carried out on Euler cluster at ETH Zurich and are openly available in the Materials Cloud Archive (www.materialscloud.org) with ID materialscloud:2019.0063 .

The authors thank Sandro Bottaro and Barak Hirshberg for useful feedback on the manuscript.
M.I. thanks Dario Azzimonti and Manuel Schuerch for drawing his attention to kernel compression algorithms.

\end{acknowledgement}

\begin{suppinfo}

Full description of the kernel compression algorithm; further details on the bandwidth rescaling; the estimate of the normalization factor $Z_n$; notes on the barrier parameter.
Detailed results of the all the test systems: double well model, alanine dipeptide, alanine tetrapeptide.
(\href{https://arxiv.org/src/1909.07250v4/anc/OPES-SI.pdf}{PDF})

\end{suppinfo}

\bibliography{biblio}

\providecommand{\latin}[1]{#1}
\makeatletter
\providecommand{\doi}
  {\begingroup\let\do\@makeother\dospecials
  \catcode`\{=1 \catcode`\}=2 \doi@aux}
\providecommand{\doi@aux}[1]{\endgroup\texttt{#1}}
\makeatother
\providecommand*\mcitethebibliography{\thebibliography}
\csname @ifundefined\endcsname{endmcitethebibliography}
  {\let\endmcitethebibliography\endthebibliography}{}
\begin{mcitethebibliography}{37}
\providecommand*\natexlab[1]{#1}
\providecommand*\mciteSetBstSublistMode[1]{}
\providecommand*\mciteSetBstMaxWidthForm[2]{}
\providecommand*\mciteBstWouldAddEndPuncttrue
  {\def\EndOfBibitem{\unskip.}}
\providecommand*\mciteBstWouldAddEndPunctfalse
  {\let\EndOfBibitem\relax}
\providecommand*\mciteSetBstMidEndSepPunct[3]{}
\providecommand*\mciteSetBstSublistLabelBeginEnd[3]{}
\providecommand*\EndOfBibitem{}
\mciteSetBstSublistMode{f}
\mciteSetBstMaxWidthForm{subitem}{(\alph{mcitesubitemcount})}
\mciteSetBstSublistLabelBeginEnd
  {\mcitemaxwidthsubitemform\space}
  {\relax}
  {\relax}

\bibitem[Peters(2017)]{Peters2017}
Peters,~B. \emph{{Reaction rate theory and rare events}}; Elsevier, 2017; p
  619\relax
\mciteBstWouldAddEndPuncttrue
\mciteSetBstMidEndSepPunct{\mcitedefaultmidpunct}
{\mcitedefaultendpunct}{\mcitedefaultseppunct}\relax
\EndOfBibitem
\bibitem[Torrie and Valleau(1977)Torrie, and Valleau]{Torrie1977}
Torrie,~G.; Valleau,~J. {Nonphysical sampling distributions in Monte Carlo
  free-energy estimation: Umbrella sampling}. \emph{Journal of Computational
  Physics} \textbf{1977}, \emph{23}, 187--199\relax
\mciteBstWouldAddEndPuncttrue
\mciteSetBstMidEndSepPunct{\mcitedefaultmidpunct}
{\mcitedefaultendpunct}{\mcitedefaultseppunct}\relax
\EndOfBibitem
\bibitem[Laio and Parrinello(2002)Laio, and Parrinello]{Laio2002}
Laio,~A.; Parrinello,~M. {Escaping free-energy minima}. \emph{Proceedings of
  the National Academy of Sciences} \textbf{2002}, \emph{99},
  12562--12566\relax
\mciteBstWouldAddEndPuncttrue
\mciteSetBstMidEndSepPunct{\mcitedefaultmidpunct}
{\mcitedefaultendpunct}{\mcitedefaultseppunct}\relax
\EndOfBibitem
\bibitem[Barducci \latin{et~al.}(2008)Barducci, Bussi, and
  Parrinello]{Barducci2008}
Barducci,~A.; Bussi,~G.; Parrinello,~M. {Well-Tempered Metadynamics: A Smoothly
  Converging and Tunable Free-Energy Method}. \emph{Physical Review Letters}
  \textbf{2008}, \emph{100}, 020603\relax
\mciteBstWouldAddEndPuncttrue
\mciteSetBstMidEndSepPunct{\mcitedefaultmidpunct}
{\mcitedefaultendpunct}{\mcitedefaultseppunct}\relax
\EndOfBibitem
\bibitem[Branduardi \latin{et~al.}(2012)Branduardi, Bussi, and
  Parrinello]{Branduardi2012}
Branduardi,~D.; Bussi,~G.; Parrinello,~M. {Metadynamics with adaptive
  gaussians}. \emph{Journal of Chemical Theory and Computation} \textbf{2012},
  \emph{8}, 2247--2254\relax
\mciteBstWouldAddEndPuncttrue
\mciteSetBstMidEndSepPunct{\mcitedefaultmidpunct}
{\mcitedefaultendpunct}{\mcitedefaultseppunct}\relax
\EndOfBibitem
\bibitem[Dama \latin{et~al.}(2014)Dama, Parrinello, and
  Voth]{Dama2014_convergence}
Dama,~J.~F.; Parrinello,~M.; Voth,~G.~A. {Well-tempered metadynamics converges
  asymptotically.} \emph{Physical review letters} \textbf{2014}, \emph{112},
  240602\relax
\mciteBstWouldAddEndPuncttrue
\mciteSetBstMidEndSepPunct{\mcitedefaultmidpunct}
{\mcitedefaultendpunct}{\mcitedefaultseppunct}\relax
\EndOfBibitem
\bibitem[Pietrucci(2017)]{Pietrucci2017}
Pietrucci,~F. {Strategies for the exploration of free energy landscapes: Unity
  in diversity and challenges ahead}. \emph{Reviews in Physics} \textbf{2017},
  \emph{2}, 32--45\relax
\mciteBstWouldAddEndPuncttrue
\mciteSetBstMidEndSepPunct{\mcitedefaultmidpunct}
{\mcitedefaultendpunct}{\mcitedefaultseppunct}\relax
\EndOfBibitem
\bibitem[Raiteri \latin{et~al.}(2006)Raiteri, Laio, Gervasio, Micheletti, and
  Parrinello]{Raiteri2006}
Raiteri,~P.; Laio,~A.; Gervasio,~F.~L.; Micheletti,~C.; Parrinello,~M.
  {Efficient Reconstruction of Complex Free Energy Landscapes by Multiple
  Walkers Metadynamics †}. \emph{The Journal of Physical Chemistry B}
  \textbf{2006}, \emph{110}, 3533--3539\relax
\mciteBstWouldAddEndPuncttrue
\mciteSetBstMidEndSepPunct{\mcitedefaultmidpunct}
{\mcitedefaultendpunct}{\mcitedefaultseppunct}\relax
\EndOfBibitem
\bibitem[Piana and Laio(2007)Piana, and Laio]{Piana2007}
Piana,~S.; Laio,~A. {A Bias-Exchange Approach to Protein Folding}. \emph{The
  Journal of Physical Chemistry B} \textbf{2007}, \emph{111}, 4553--4559\relax
\mciteBstWouldAddEndPuncttrue
\mciteSetBstMidEndSepPunct{\mcitedefaultmidpunct}
{\mcitedefaultendpunct}{\mcitedefaultseppunct}\relax
\EndOfBibitem
\bibitem[Dama \latin{et~al.}(2014)Dama, Rotskoff, Parrinello, and
  Voth]{Dama2014_ttmetad}
Dama,~J.~F.; Rotskoff,~G.; Parrinello,~M.; Voth,~G.~A. {Transition-tempered
  metadynamics: Robust, convergent metadynamics via on-the-fly transition
  barrier estimation}. \emph{Journal of Chemical Theory and Computation}
  \textbf{2014}, \emph{10}, 3626--3633\relax
\mciteBstWouldAddEndPuncttrue
\mciteSetBstMidEndSepPunct{\mcitedefaultmidpunct}
{\mcitedefaultendpunct}{\mcitedefaultseppunct}\relax
\EndOfBibitem
\bibitem[White \latin{et~al.}(2015)White, Dama, and Voth]{White2015}
White,~A.~D.; Dama,~J.~F.; Voth,~G.~A. {Designing Free Energy Surfaces That
  Match Experimental Data with Metadynamics}. \emph{Journal of Chemical Theory
  and Computation} \textbf{2015}, \emph{11}, 2451--2460\relax
\mciteBstWouldAddEndPuncttrue
\mciteSetBstMidEndSepPunct{\mcitedefaultmidpunct}
{\mcitedefaultendpunct}{\mcitedefaultseppunct}\relax
\EndOfBibitem
\bibitem[Pfaendtner and Bonomi(2015)Pfaendtner, and Bonomi]{Pfaendtner2015}
Pfaendtner,~J.; Bonomi,~M. {Efficient Sampling of High-Dimensional Free-Energy
  Landscapes with Parallel Bias Metadynamics}. \emph{Journal of Chemical Theory
  and Computation} \textbf{2015}, \emph{11}, 5062--5067\relax
\mciteBstWouldAddEndPuncttrue
\mciteSetBstMidEndSepPunct{\mcitedefaultmidpunct}
{\mcitedefaultendpunct}{\mcitedefaultseppunct}\relax
\EndOfBibitem
\bibitem[Bonomi \latin{et~al.}(2009)Bonomi, Barducci, and
  Parrinello]{Bonomi2009}
Bonomi,~M.; Barducci,~A.; Parrinello,~M. {Reconstructing the equilibrium
  Boltzmann distribution from well-tempered metadynamics}. \emph{Journal of
  Computational Chemistry} \textbf{2009}, \emph{30}, 1615--1621\relax
\mciteBstWouldAddEndPuncttrue
\mciteSetBstMidEndSepPunct{\mcitedefaultmidpunct}
{\mcitedefaultendpunct}{\mcitedefaultseppunct}\relax
\EndOfBibitem
\bibitem[Tiwary and Parrinello(2015)Tiwary, and Parrinello]{Tiwary2015}
Tiwary,~P.; Parrinello,~M. {A time-independent free energy estimator for
  metadynamics}. \emph{Journal of Physical Chemistry B} \textbf{2015},
  \emph{119}, 736--742\relax
\mciteBstWouldAddEndPuncttrue
\mciteSetBstMidEndSepPunct{\mcitedefaultmidpunct}
{\mcitedefaultendpunct}{\mcitedefaultseppunct}\relax
\EndOfBibitem
\bibitem[Mones \latin{et~al.}(2016)Mones, Bernstein, and
  Cs{\'{a}}nyi]{Mones2016}
Mones,~L.; Bernstein,~N.; Cs{\'{a}}nyi,~G. {Exploration, Sampling, And
  Reconstruction of Free Energy Surfaces with Gaussian Process Regression}.
  \emph{Journal of Chemical Theory and Computation} \textbf{2016}, \emph{12},
  5100--5110\relax
\mciteBstWouldAddEndPuncttrue
\mciteSetBstMidEndSepPunct{\mcitedefaultmidpunct}
{\mcitedefaultendpunct}{\mcitedefaultseppunct}\relax
\EndOfBibitem
\bibitem[Donati and Keller(2018)Donati, and Keller]{Donati2018}
Donati,~L.; Keller,~B.~G. {Girsanov reweighting for metadynamics simulations}.
  \emph{The Journal of Chemical Physics} \textbf{2018}, \emph{149},
  072335\relax
\mciteBstWouldAddEndPuncttrue
\mciteSetBstMidEndSepPunct{\mcitedefaultmidpunct}
{\mcitedefaultendpunct}{\mcitedefaultseppunct}\relax
\EndOfBibitem
\bibitem[Marinova and Salvalaglio(2019)Marinova, and
  Salvalaglio]{Salvalaglio2019}
Marinova,~V.; Salvalaglio,~M. {Time-independent free energies from metadynamics
  via mean force integration}. \emph{The Journal of Chemical Physics}
  \textbf{2019}, \emph{151}, 164115\relax
\mciteBstWouldAddEndPuncttrue
\mciteSetBstMidEndSepPunct{\mcitedefaultmidpunct}
{\mcitedefaultendpunct}{\mcitedefaultseppunct}\relax
\EndOfBibitem
\bibitem[Giberti \latin{et~al.}(2019)Giberti, Cheng, Tribello, and
  Ceriotti]{Giberti2019}
Giberti,~F.; Cheng,~B.; Tribello,~G.~A.; Ceriotti,~M. {Iterative unbiasing of
  quasi-equilibrium sampling}. \textbf{2019}, \relax
\mciteBstWouldAddEndPunctfalse
\mciteSetBstMidEndSepPunct{\mcitedefaultmidpunct}
{}{\mcitedefaultseppunct}\relax
\EndOfBibitem
\bibitem[Invernizzi \latin{et~al.}(2017)Invernizzi, Valsson, and
  Parrinello]{Invernizzi2017}
Invernizzi,~M.; Valsson,~O.; Parrinello,~M. {Coarse graining from variationally
  enhanced sampling applied to the Ginzburg–Landau model}. \emph{Proceedings
  of the National Academy of Sciences} \textbf{2017}, \emph{114},
  3370--3374\relax
\mciteBstWouldAddEndPuncttrue
\mciteSetBstMidEndSepPunct{\mcitedefaultmidpunct}
{\mcitedefaultendpunct}{\mcitedefaultseppunct}\relax
\EndOfBibitem
\bibitem[Mezei(1987)]{Mezei1987}
Mezei,~M. {Adaptive umbrella sampling: Self-consistent determination of the
  non-Boltzmann bias}. \emph{Journal of Computational Physics} \textbf{1987},
  \emph{68}, 237--248\relax
\mciteBstWouldAddEndPuncttrue
\mciteSetBstMidEndSepPunct{\mcitedefaultmidpunct}
{\mcitedefaultendpunct}{\mcitedefaultseppunct}\relax
\EndOfBibitem
\bibitem[Maragakis \latin{et~al.}(2009)Maragakis, van~der Vaart, and
  Karplus]{Maragakis2009}
Maragakis,~P.; van~der Vaart,~A.; Karplus,~M. {Gaussian-Mixture Umbrella
  Sampling}. \emph{The Journal of Physical Chemistry B} \textbf{2009},
  \emph{113}, 4664--4673\relax
\mciteBstWouldAddEndPuncttrue
\mciteSetBstMidEndSepPunct{\mcitedefaultmidpunct}
{\mcitedefaultendpunct}{\mcitedefaultseppunct}\relax
\EndOfBibitem
\bibitem[Marsili \latin{et~al.}(2006)Marsili, Barducci, Chelli, Procacci, and
  Schettino]{Marsili2006}
Marsili,~S.; Barducci,~A.; Chelli,~R.; Procacci,~P.; Schettino,~V.
  {Self-healing Umbrella Sampling: A Non-equilibrium Approach for Quantitative
  Free Energy Calculations}. \emph{The Journal of Physical Chemistry B}
  \textbf{2006}, \emph{110}, 14011--14013\relax
\mciteBstWouldAddEndPuncttrue
\mciteSetBstMidEndSepPunct{\mcitedefaultmidpunct}
{\mcitedefaultendpunct}{\mcitedefaultseppunct}\relax
\EndOfBibitem
\bibitem[Dickson \latin{et~al.}(2010)Dickson, Legoll, Leli{\`{e}}vre, Stoltz,
  and Fleurat-Lessard]{Dickson2010}
Dickson,~B.~M.; Legoll,~F.; Leli{\`{e}}vre,~T.; Stoltz,~G.; Fleurat-Lessard,~P.
  {Free Energy Calculations: An Efficient Adaptive Biasing Potential Method}.
  \emph{The Journal of Physical Chemistry B} \textbf{2010}, \emph{114},
  5823--5830\relax
\mciteBstWouldAddEndPuncttrue
\mciteSetBstMidEndSepPunct{\mcitedefaultmidpunct}
{\mcitedefaultendpunct}{\mcitedefaultseppunct}\relax
\EndOfBibitem
\bibitem[Valsson and Parrinello(2014)Valsson, and Parrinello]{Valsson2014}
Valsson,~O.; Parrinello,~M. {Variational approach to enhanced sampling and free
  energy calculations}. \emph{Physical Review Letters} \textbf{2014},
  \emph{113}, 1--5\relax
\mciteBstWouldAddEndPuncttrue
\mciteSetBstMidEndSepPunct{\mcitedefaultmidpunct}
{\mcitedefaultendpunct}{\mcitedefaultseppunct}\relax
\EndOfBibitem
\bibitem[Valsson and Parrinello(2015)Valsson, and Parrinello]{Valsson2015}
Valsson,~O.; Parrinello,~M. {Well-tempered variational approach to enhanced
  sampling}. \emph{Journal of Chemical Theory and Computation} \textbf{2015},
  \emph{11}, 1996--2002\relax
\mciteBstWouldAddEndPuncttrue
\mciteSetBstMidEndSepPunct{\mcitedefaultmidpunct}
{\mcitedefaultendpunct}{\mcitedefaultseppunct}\relax
\EndOfBibitem
\bibitem[Shaffer \latin{et~al.}(2016)Shaffer, Valsson, and
  Parrinello]{Shaffer2016}
Shaffer,~P.; Valsson,~O.; Parrinello,~M. {Enhanced, targeted sampling of
  high-dimensional free-energy landscapes using variationally enhanced
  sampling, with an application to chignolin}. \emph{Proceedings of the
  National Academy of Sciences} \textbf{2016}, \emph{113}, 1150--1155\relax
\mciteBstWouldAddEndPuncttrue
\mciteSetBstMidEndSepPunct{\mcitedefaultmidpunct}
{\mcitedefaultendpunct}{\mcitedefaultseppunct}\relax
\EndOfBibitem
\bibitem[Debnath \latin{et~al.}(2019)Debnath, Invernizzi, and
  Parrinello]{Debnath2019}
Debnath,~J.; Invernizzi,~M.; Parrinello,~M. {Enhanced Sampling of Transition
  States}. \emph{Journal of Chemical Theory and Computation} \textbf{2019},
  \emph{15}, 2454--2459\relax
\mciteBstWouldAddEndPuncttrue
\mciteSetBstMidEndSepPunct{\mcitedefaultmidpunct}
{\mcitedefaultendpunct}{\mcitedefaultseppunct}\relax
\EndOfBibitem
\bibitem[Piaggi and Parrinello(2019)Piaggi, and Parrinello]{Piaggi2019}
Piaggi,~P.~M.; Parrinello,~M. {Multithermal-Multibaric Molecular Simulations
  from a Variational Principle}. \emph{Physical Review Letters} \textbf{2019},
  \emph{122}, 050601\relax
\mciteBstWouldAddEndPuncttrue
\mciteSetBstMidEndSepPunct{\mcitedefaultmidpunct}
{\mcitedefaultendpunct}{\mcitedefaultseppunct}\relax
\EndOfBibitem
\bibitem[Invernizzi and Parrinello(2019)Invernizzi, and
  Parrinello]{Invernizzi2019}
Invernizzi,~M.; Parrinello,~M. {Making the Best of a Bad Situation: A
  Multiscale Approach to Free Energy Calculation}. \emph{Journal of Chemical
  Theory and Computation} \textbf{2019}, \emph{15}, 2187--2194\relax
\mciteBstWouldAddEndPuncttrue
\mciteSetBstMidEndSepPunct{\mcitedefaultmidpunct}
{\mcitedefaultendpunct}{\mcitedefaultseppunct}\relax
\EndOfBibitem
\bibitem[Silverman(1998)]{Silverman1998}
Silverman,~B. \emph{{Density Estimation for Statistics and Data Analysis}};
  Routledge: New York, 1998\relax
\mciteBstWouldAddEndPuncttrue
\mciteSetBstMidEndSepPunct{\mcitedefaultmidpunct}
{\mcitedefaultendpunct}{\mcitedefaultseppunct}\relax
\EndOfBibitem
\bibitem[Kish(1965)]{Kish1965}
Kish,~L. \emph{{Sampling Organizations and Groups of Unequal Sizes}}; 1965;
  Vol.~30; pp 564--572\relax
\mciteBstWouldAddEndPuncttrue
\mciteSetBstMidEndSepPunct{\mcitedefaultmidpunct}
{\mcitedefaultendpunct}{\mcitedefaultseppunct}\relax
\EndOfBibitem
\bibitem[Sodkomkham \latin{et~al.}(2016)Sodkomkham, Ciliberti, Wilson, Fukui,
  Moriyama, Numao, and Kloosterman]{Sodkomkham2016}
Sodkomkham,~D.; Ciliberti,~D.; Wilson,~M.~A.; Fukui,~K.-I.; Moriyama,~K.;
  Numao,~M.; Kloosterman,~F. {Kernel density compression for real-time Bayesian
  encoding/decoding of unsorted hippocampal spikes}. \emph{Knowledge-Based
  Systems} \textbf{2016}, \emph{94}, 1--12\relax
\mciteBstWouldAddEndPuncttrue
\mciteSetBstMidEndSepPunct{\mcitedefaultmidpunct}
{\mcitedefaultendpunct}{\mcitedefaultseppunct}\relax
\EndOfBibitem
\bibitem[Tribello \latin{et~al.}(2014)Tribello, Bonomi, Branduardi, Camilloni,
  and Bussi]{plumed}
Tribello,~G.~A.; Bonomi,~M.; Branduardi,~D.; Camilloni,~C.; Bussi,~G. {PLUMED
  2: New feathers for an old bird}. \emph{Computer Physics Communications}
  \textbf{2014}, \emph{185}, 604--613\relax
\mciteBstWouldAddEndPuncttrue
\mciteSetBstMidEndSepPunct{\mcitedefaultmidpunct}
{\mcitedefaultendpunct}{\mcitedefaultseppunct}\relax
\EndOfBibitem
\bibitem[Fort \latin{et~al.}(2017)Fort, Jourdain, Leli{\`{e}}vre, and
  Stoltz]{Fort2017}
Fort,~G.; Jourdain,~B.; Leli{\`{e}}vre,~T.; Stoltz,~G. {Self-healing umbrella
  sampling: convergence and efficiency}. \emph{Statistics and Computing}
  \textbf{2017}, \emph{27}, 147--168\relax
\mciteBstWouldAddEndPuncttrue
\mciteSetBstMidEndSepPunct{\mcitedefaultmidpunct}
{\mcitedefaultendpunct}{\mcitedefaultseppunct}\relax
\EndOfBibitem
\bibitem[{The PLUMED consortium}(2019)]{Plumed2019}
{The PLUMED consortium}, {Promoting transparency and reproducibility in
  enhanced molecular simulations}. \emph{Nature Methods} \textbf{2019},
  \emph{16}, 670--673\relax
\mciteBstWouldAddEndPuncttrue
\mciteSetBstMidEndSepPunct{\mcitedefaultmidpunct}
{\mcitedefaultendpunct}{\mcitedefaultseppunct}\relax
\EndOfBibitem
\bibitem[Leli{\`{e}}vre \latin{et~al.}(2010)Leli{\`{e}}vre, Rousset, and
  Stoltz]{Lelievre2010}
Leli{\`{e}}vre,~T.; Rousset,~M.; Stoltz,~G. \emph{{Free Energy Computations}};
  IMPERIAL COLLEGE PRESS, 2010\relax
\mciteBstWouldAddEndPuncttrue
\mciteSetBstMidEndSepPunct{\mcitedefaultmidpunct}
{\mcitedefaultendpunct}{\mcitedefaultseppunct}\relax
\EndOfBibitem
\end{mcitethebibliography}

\end{document}